\documentstyle[11pt,newpasp,twoside,epsf]{article}
\def\edcomment#1{\iffalse\marginpar{\raggedright\sl#1\/}\else\relax\fi}
\reversemarginpar

\begin{document}
\title{Pulsar Searches of Unidentified $EGRET$ Sources}
 \author{Mallory S.E. Roberts\altaffilmark{1}}
\affil{McGill University, 3600 University St. Montr\'eal, QC. H3A 2T8, Canada
}
\altaffiltext{1}{Massachusetts Institute of Technology, Cambridge, MA, USA}

\begin{abstract}
The majority of Galactic high-energy $\gamma$-ray sources continue
to elude identification. Currently, we have a handful of firm pulsar
identifications, one of which is radio quiet, and a few marginal 
detections, including one millisecond pulsar. Recently, both blind 
searches of $EGRET$ error boxes and targeted searches of X-ray counterpart
candidates have had some success in finding new pulsars. I review these results,
and discuss our current program of searching mid-Galactic latitude 
$EGRET$ error boxes using the Parkes multi-beam system.  
\end{abstract}

\section{Introduction}

The $EGRET$ $\gamma$-ray telescope detected around 300 sources at
energies above 100 MeV, the majority of which remain unidentified
(Hartman et al. 1999). 
Large error boxes on the order of $1^{\circ}$ across make 
unambiguous identifications on the basis of source position
difficult. Timing observations allow definitive identifications
for sources showing aperiodic (such as active galactic nuclei) 
or periodic (pulsars) variability. Between 6-8 pulsars have been 
positively identified in the $EGRET$ data by epoch folding on the
known period determined from radio or, in the singular case of Geminga, 
X-ray data (Thompson et al 1999, Kaspi et al. 2000 and references therein).
Several other energetic pulsars were
found to be coincident with unidentified $EGRET$ sources, but 
folding the sparse $\gamma$-ray data 
using contemporaneous radio ephemerides did not yield 
significant detections (Fierro, 1995).
Although $GLAST$, when it flies in 2006, may be able
to discover new pulsars  just by searching the $\gamma$-ray data, 
it is likely that
a priori knowledge of a pulse period will in many cases prove crucial in 
determining which $\gamma$-ray sources are pulsars.

\section{Recent Galactic Plane Pulsar Discoveries}

Since the effective end of the $EGRET$ mission and the publishing 
of the third $EGRET$ catalog (Hartman et al. 1999), there have been both
new wide scale pulsar surveys and searches of potential $\gamma$-ray
counterparts which have yielded several new, energetic young pulsars
positionally coincident with unidentified $EGRET$ sources. The most
productive effort has been the Parkes multibeam survey (Manchester et al. 2001).
Manchester et al. (2002) list six energetic pulsars from the PM survey which
are contained within $EGRET$ positional error boxes. For two of these,
PSR J1016$-$5857 and PSR J1420$-$6048, previously
obtained X-ray and radio data identified potential 
pulsar wind nebulae (PWN) (Camilo et al. 2001, Roberts, Romani, \& Johnston 
2001). Hence, 
searches directed at the potential counterparts would have been successful.
In a third case, PSR J1015$-$5719, very little is known about any 
potential X-ray or radio PWN. In the last three cases, neither radio nor
X-ray data show clear evidence for a potential source counterpart.
PSR J1837$-$0604 is on the edge of a bright molecular cloud, making 
detection of a radio PWN difficult (D'Amico et al. 2001). A 10 ks
{\it Chandra} image of the region shows no evidence for a X-ray point source 
associated with the pulsar to extremely low flux levels (Roberts et al. 
in preparation). 
The last two Parkes multibeam pulsars, PSR J1413$-$6141 and PSR J1412$-$6145,
are located within the bright energetic supernova remnant G312.4$-$0.4. 
{\it Chandra} imaging and radio spectral tomography of the region show
neither point sources nor potential PWN associated with either pulsar
(Doherty et al. in press). The low spin-down luminosity, large distance, and
alternate potential source of $\gamma$-rays (the supernova remnant) all
argue against either of these latter two pulsars being the true
counterpart of the coincident $EGRET$ source.   

Deep pulse searches of potential X-ray and radio counterparts have also
recently proven successful. The extremely energetic pulsar 
PSR J2229+6114 was discovered with the Lovell telescope at 
Jodrell Bank in a 1.5~hr observation of a potential 
PWN identified by X-ray and radio imaging of the associated
$EGRET$ error box (Halpern et al. 2001). An $ASCA$ X-ray image of
the unidentified source GeV J2020+3658, associated with the
COS B source 2CG 075+00, showed
a point-like X-ray source at which the very small ($3^{\prime}$)
Arecibo beam could be directed. The young pulsar PSR J2021+3651, at
a 20~cm flux of only 0.1 mJy, was discovered as a result (Roberts et al. 2002). 
This flux is about a factor of two 
below the sensitivity limits of the Parkes multibeam 
survey and similar to the sensitivity limit of a 5~hr observation
with Parkes. Thus, it is possible that deep searches of potential 
counterparts contained within the Parkes Multibeam survey region 
will yield further candidates. However, three likely PWN identified by
Roberts, Romani, and Kawai (2001) were searched with 5~hr 
pointings at Parkes and failed to yield any good candidates. Two 
of these, the Rabbit and GeV J1809$-$2327, are almost certainly 
PWN, based on their X-ray and radio properties (Roberts et al. 1999,
Braje et al. 2002). Therefore, more sensitive radio searches might 
be necessary to detect the pulsars these contain. The third source,
GeV J1825-1310 is less clear of a PWN candidate, due to its large size
and the confused radio emission in the region, but other 
source classes for hard, extended X-ray sources seem less plausible.

Between the Parkes Multibeam survey and directed searches, six 
new strong $\gamma$-ray pulsar candidates have been discovered as
radio pulsars since the demise of $EGRET$. Since these are young pulsars,
which tend to glitch, extrapolating ephemerides back to the $EGRET$
era is uncertain making detection of $\gamma$-ray pulsations difficult.
I list these pulsars in Table 1 in boldface.
The dispersion measure (DM) distances, determined 
using the new Cordes and Lazio (2002) NE2001 model, are given.
The $\gamma$-ray
efficiencies, $\eta_{\gamma}$, are determined from the 100~MeV to
10~GeV flux, assuming the DM distance and 1 steradian beaming.
All of the new pulsars have
high DMs and have 20~cm fluxes $< 1$ mJy. The inferred $\gamma$-ray 
efficiencies are also quite high. For comparison, Table 1 lists the 
values for the known $\gamma$-ray pulsars, 
using the NE2001 distances for consistency and  
adapting the efficiencies from the values given by Kaspi et al. (2000).
If the $\gamma$-ray sources associated with these new pulsars are
nearly 100\% pulsed, as are the known sources, 
then the oft-noted inverse correlation of efficiency
with spin-down power ($\eta_{\gamma} \propto L_{sd}^{-1/2}$,  eg. Zhang \& Harding 2000) 
and characteristic age, $\tau_c$, is not well supported. 

\begin{table}
\caption{Radio/$\gamma$-ray? Pulsars} 
\begin{tabular}{lcccccc}
\tableline
Pulsar & log $\dot E$ & log $\tau_c$ & DM & $D_{NE2001}$ & $S_{20cm}$ &
$\eta_{\gamma}$ \\  
& ${\rm erg}\, {\rm cm}^{-2}{\rm s}^{-1}$ &yr& pc$\,{\rm cm}^{-3}$ & kpc & mJy & \\
\tableline
Crab & 38.7 & 3.1 & 57 & 1.7 & 14 & 0.0001 \\
{\bf J2229+6114} & 37.3 & 4.0 & 200 & 7.3 & 0.3 & 0.007 \\
{\bf J1420$-$6048} & 37.0 & 4.1 & 360 & 5.6 & 0.9 & 0.01 \\
Vela & 36.8 & 4.1 & 68 & 0.24 & 1000 & 0.0004 \\
B1951+32 & 36.6 & 5.0 & 45 & 3.2 & 1.0 & 0.004 \\
B1706$-$44 & 36.5 & 4.2 & 76 & 2.3 & 10 & 0.01 \\
{\bf J2021+3651} & 36.5 & 4.2 & 369 & 12 & 0.1 & 0.2 \\
{\bf J1016$-$5857} & 36.4 & 4.3 & 394 & 8.0 & 0.5 & 0.03 \\
B1046$-$58 & 36.3 & 4.3 & 129 & 2.7 & 8 & 0.009 \\
{\bf J1837$-$0604} & 36.3 & 4.5 & 462 & 6.4 & 0.4 & 0.07 \\
{\bf J1015$-$5719} & 35.9 & 4.6 & 283 & 5.1 & 0.9 & 0.1 \\
B0656+14 & 34.6 & 5.0 & 14 & 0.67 & 3.7 & 0.0007 \\
Geminga & 34.5 & 5.5 & - & 0.16\tablenotemark{a} & - & 0.02 \\
B1055-52 & 34.5 & 5.7 & 30 & 0.72 & - & 0.03 \\
\tableline
\tableline
\end{tabular}
\footnotesize{a.Not a radio pulsar, no DM distance available}
\end{table}

\section{A Mid-Latitude Survey for $\gamma$-ray Pulsars}

In addition to the Galactic plane population, there is at least one
distinct population of unidentified $\gamma$-ray sources at mid-Galactic latitudes.
Some authors have modelled these as a Gould belt (a local region 
of recent star formation) population plus a Galactic Halo population
(Grenier 2001). It has also been noted that the distribution is similar to
that of the millisecond pulsars (Romani 2001). A population of older, 
Geminga-like pulsars has been postulated, as well as a population of
$\gamma$-ray emitting millisecond pulsars. Indeed, there has been a 
recent claim of a marginal detection of pulsed $\gamma$-rays in the 
$EGRET$ data set from the millisecond pulsar PSR J0218+4232
(Kuiper et al. 2000). Practically speaking, because the mid-latitude
sources are generally weaker and have steeper spectra than 
the Galactic plane sources, they will be difficult to identify 
if they are not pulsed or have some other characteristic variability. 
If some fraction are pulsars, then a priori knowledge of the pulse
period may be crucial to finding the pulsations at $\gamma$-ray 
energies even with the large effective area of the $GLAST$ mission. 

We have begun a survey of the unidentified mid-latitude $EGRET$ sources
that are observable by the Parkes telescope. With the multibeam system, 
a series of four pointings in the standard tesselation pattern can 
completely cover most $\gamma$-ray error boxes out to the 95\% confidence
contour. Alist of 56 sources from the 3rd $EGRET$ catalog at latitudes 
$|b| > 5\deg$ have been selected on the basis of their error box size
and lack of probable AGN counterpart. Four 35~min pointings using a time
resolution of 0.125~ms cover each of the error boxes. In order to
be sensitive to millisecond pulsars in tight binaries, a full acceleration
search will be performed using the {\tt PRESTO} software package (Ransom 2001).
The sensitivity of this survey will be similar to that of the PM 
survey, and much greater than the highly successful
mid-latitude surveys with Parkes, which generally use only 4 min. pointings
(Edwards et al. 2001).

\end{document}